\documentclass[11pt, reqno]{amsart}
\usepackage{graphicx}
\usepackage{amsfonts}
\usepackage{amsmath}
\usepackage{float}
\usepackage{color}
\usepackage[english]{babel}

\newtheorem{theorem}{Theorem}[section]
\newtheorem{lemma}{Lemma}[section]
\newtheorem{corollary}[theorem]{Corollary}

\newtheorem{remark}{Remark}[section]
\newcommand{\RR}{\mathbb{R}}
\newcommand{\CC}{\mathbb{C}}

\newcommand{\R}{{\mathbb R}}

\def\rem2{\noindent   {\bf Remark.} }

\numberwithin{equation}{section}

\def\uomega{{\underline \omega}}

\def\uomega{{\underline \omega}}

\newcommand{\eps}{\varepsilon}
\renewcommand{\phi}{\varphi}

\def\bfe{{\bf e}}
\def\S^2{{\mathbb S}^2}

\def\xx{\frac{x}{|x|}}

\def\bfw{{\bf w}}

\def\bfe{{\bf e}}
\def\bfb{{\bf b}}

\def\caD{{\mathcal D}}
\def\caN{{\mathcal N}}

\def\R{{\mathbb R}}

\def\curl{{\rm curl}\,}
\def\dive{{\rm div}\,}
\def\grad{{\rm grad}\,}
\def\tn{{\rm tan}\,}

\numberwithin{equation}{section}

\title{Incoming  and Disappearing Solutions for Maxwell's Equations}

\author[F. Colombini, V. Petkov] {Ferruccio Colombini, \ Vesselin Petkov\ }
\address{Dipartimento di Matematica, Universit\`a di Pisa, Italia}
\email{colombini@dm.unipi.it}
\address{Institut de Math\'ematiques de Bordeaux, 351,
Cours de la Lib\'eration, 33405  Talence, France}
\email{petkov@math.u-bordeaux1.fr}
\author[J. Rauch]{\ Jeffrey Rauch$\,^\dagger$}
\thanks{$^\dagger$ Research partially supported by the National Science
Foundation under grant  NSF DMS 0405899}
\address{Department of Mathematics, University of Michigan, USA}
\email{rauch@umich.edu}

\numberwithin{equation}{section}

\date{}

\begin{document}
\maketitle

\begin{abstract} We prove that in contrast to the free wave equation in $\R^3$ there are no incoming solutions of Maxwell's equations in the form of 
spherical or modulated spherical waves.
We construct solutions which are corrected by 
lower order incoming waves.  With their aid, we
construct dissipative boundary conditions and solutions
to Maxwell's equations in the exterior of a sphere which
decay exponentially 
 as $t \to +\infty$. They are asymptotically disappearing. 
Disappearing solutions which are identically zero for 
$t \geq T > 0$ are constructed which
satisfy maximal dissipative boundary conditions which depend on time $t$. 
Both types are invisible in scattering theory.
\end{abstract}

$\:\:\:\:\:\:\:$Key words: Maxwell equations, disappearing solutions, dissipative boundary conditions\\

$\:\:\:\:\:\:\:\:$MSC2010: Primary 35Q61, Secondary 35P25, 35L45\\

\section{Introduction.}

This paper is devoted to the construction of incoming
solutions of Maxwell's equations in the exterior of a sphere
which are totally absorbed at the boundary where
a boundary condition defining a well posed initial boundary
value problem is  imposed.  By definition,
 {\it disappearing solutions} $u(t, x)$ vanish for $t \geq T_0 > 0$ and all $x$ outside a bounded obstacle. 
We also construct solutions whose  total energy 
 decays exponentially
as $t \to +\infty$.  We call such solutions
 {\it asymptotically disappearing}.
 For both types of solutions
   the scattered field vanishes identically.
    They are also  of interest from
the point of view of inverse problems as the information in the
incoming wave is irretrievably lost (see \cite{M} and
Chapter IV in \cite{P} for more comments).

There are some classic examples.  
The first one is the equation
$$
\partial_t u \ -\ \partial_x u \ =\ 0,
\quad{\rm in}\ \  x> 0,
$$
with no boundary condition imposed at $x=0$.   Compactly supported
data disappear in finite time and all solutions 
with square integrable initial data tend to zero in 
$L^2(\{x>0\})$ 
as $t \to \infty$.

The wave equation in dimension $d =1$ with absorbing boundary condition is
$$
\partial^2_t u\ -\ \partial^2_x
u\ =\ 0
\quad
{\rm in} \ \ x>0,\qquad
\partial_t u(t,0)  - \partial_x u(t,0) \ =\ 0.
$$
The standard energy is decreasing in time and
leftward moving waves 
which satisfy $\partial_tu-\partial_x u=0 $
are absorbed at the boundary without reflection.

In these two cases a significant fraction of solutions
disappear.  The situation is radically different 
for the wave equation in dimensions
$d>1$ where it is not easy to produce a single example
that disappears, that is to find solution $u(t, x)$ which vanishes for $t \geq T_0 > 0$.  The classic example is in the fundamental paper of  Majda \cite{M}.
He considers for spatial dimension $d=3$ the differential equation and
boundary condition
$$
\partial^2_t
u -\ \Delta u \ =\ 0
\quad
{\rm in}\ \ 
r:=
|x|\ge R,
\qquad
(\partial_t -\partial_r)(ru) \ =\ 0
\quad{\rm on}\ \ |x|=R\,.
$$
All incoming spherical waves 
$$
u (t, x) = \frac{f( |x| + t)}{|x|}
$$
 satisfy the boundary condition $(\partial_t -\partial_r)(ru) = 0$. 
Choosing $f \in C_0^{\infty}(\R)$ yields a disappearing solution.
We observe that the natural absorbing
boundary condition $\partial_t u - \partial_r u = 0$ is satisfied modulo a correction $-{f(|x| + t)}/{|x|^2}$.
The modified  boundary condition is satisfied exactly. 
A similar phenomenon will be described 
for Maxwell's system in Section 3.

For Maxwell's equations
\begin{equation} 
\label{eq:1.1}
 E_t = \curl B, \qquad  B_t = - \curl  E, 
\end{equation}
\begin{equation} 
\label{eq:1.2}
\dive E \ =\  \dive B \ =\  0\,,
\end{equation}
 Georgiev \cite{G} constructed, using spherical coordinates,
disappearing fields which are spherical waves modulated in the 
angular direction, that is $E$ and $B$ are of the form
$$
\frac{f(|x|-t\,,\, x/|x|)}
{|x|}\,.
$$
His solution is singular along the $z$ axis $\{x = y =0\}$, where it is locally integrable so defines a well 
defined distribution. The fields satisfy Maxwell's
equations on the complement of the $z$ axis.  We prove below that
they are not solutions 
in a neighborhood of this axis.   In Section 2 we prove 
that  Maxwell's equations have no solutions 
with the structure of modulated spherical waves.  In 
Section 3 we construct 
incoming
solutions for Maxwell's
equations which have similar form, namely,
  $$
  \frac{f_1(t+|x|\,,\,x/|x|)}{|x|}
  \ +\
   \frac{f_2(t+|x|\,,\,x/|x|)}{|x|^2}
  \ +\
   \frac{f_3(t+|x|\,,\,x/|x|)}{|x|^3}\,.
$$

As for the wave equation, the 
natural absorbing condition
$$
E_{\tn} - n(x) \wedge B_{\tn} = 0, \qquad |x| = 1,
$$
$n(x)$ being the unit normal to $|x| = 1$ pointing into $|x| \leq 1$
is not satisfied.
 In Theorem \ref{thm:asympt} we construct
asymptotically disappearing solutions satisfying a small perturbation of this maximal dissipative boundary condition.
  With small {\it and time dependent}
perturbations we construct 
disappearing solutions
in Theorem \ref{thm:disapp}.
The problem of constructing a disappearing 
solution of a time independent dissipative boundary
value problem for Maxwell's equations remains open.

\section{No modulated spherical Maxwell
solutions.}
 \label{sec:nogo}

The large time asymptotics for solutions
of the wave equation on $\R^{1+3}$ 
with Cauchy data in the Schwartz space
${\mathcal S}(\R^3)$ is  given in terms of a profile $h(s,\omega)\in {\mathcal S}(\R \times \S^2)$
as
$$ u \ =\ 
\frac{h(|x|-t\,,\,x/|x|)}{|x|}  \
+ {\mathcal O}(|x|^{-2})\,.$$
The expression
$h( |x|-t\,,\,\omega)/|x|$ is for each $\omega$ an outgoing spherical wave
solution of the wave equation in $\{x\ne 0\}$.
 
Consider next
Maxwell's equations (\ref{eq:1.1}),\: (\ref{eq:1.2}).
They imply
\begin{equation}
\label{eq:2.1}
\Box E  \ =\ \Box B\ =\ 0\,.
\end{equation}
For Cauchy data in ${\mathcal S}$ the asymptotics are given
in terms of profiles
\begin{equation}
\label{eq:edot}
\bfe(s,\omega)\quad {\rm satisfying}
\quad
\omega.\bfe=0\,,
\qquad
\bfb(x,\omega)
\ :=\
\omega\wedge \bfe\,.
\end{equation}
As $t\to +\infty$,
$$
\big(E(t,x),B(t,x)\big)=
\bigg(\frac{\bfe(|x|-t\,,\,x/|x|)}{|x|} \,,\,
\frac{\bfb(|x|-t\,,\,x/|x|)}{|x|}\bigg)+
{\mathcal O}\:\bigg(\frac{1}{t^2+|x|^2}
\bigg).
$$

The second condition $\omega.\bfe=0$ asserts that for
each $s$, $\omega\mapsto \bfe(s,\omega)$ is a tangent vector field on $\S^2$.   
Thus as soon as it is continuous, Brouwer's Theorem
implies that there is at least one $\uomega\in \S^2$
where $\bfe(s,\uomega)=\bfb(s,\uomega)=0$.  There are
always quiet spots in the far field.

In analogy with the spherical wave solutions of the wave
equation it is natural to ask if there are some special
profiles so that
$$
(E,B)
\ :=\ \bigg(\frac{\bfe(|x|-t\,,\,x/|x|)}{|x|}\,,\,
\frac{\bfb(|x|-t\,,\,x/|x|)}{|x|} \bigg)
$$
exactly satisfies Maxwell's equations in $\{x\ne 0\}$.
With
$$
w(t,r,\omega)
\ :=\
\bigg(\frac{\bfe(r-t\,,\,\omega)}{r}\,,\,
\frac{\bfb(r-t\,,\,\omega)}{r}
\bigg)
$$
 this is an  expression of the form
\begin{equation}
\label{eq:ansatz}
w(t,|x|, x/|x|),
\qquad
{\rm where}
\qquad
\Box_{t,x}w(t,|x|, \omega)\ =\ 0\,.
\end{equation}

 We next prove that it is impossible for a fields as in
 \eqref{eq:ansatz}
 to satisfy Maxwell's equations unless they are independent
 of $x$.
 Thus,
 \vskip.1cm

\noindent 
 $\bullet$  The leading expression in the asymptotic description
 is never an exact solution.
 
 \noindent $\bullet$  There is no natural analogue for Maxwell's equations
 of exact
 spherical wave solutions of the wave equation. 
  
 \noindent $\bullet$   Georgiev's disappearing
 solutions \cite{G} do not satisfy Maxwell's equations on the $z$-axis.
  
 \vskip.1cm
 
 Suppose that $I$ is a nontrivial open  interval
in $\RR$, $0<a<b\le \infty,$  and $v$ is a distribution on 
$I\times]a,b[\times \S^2$.  If $v$ is smooth and 
$\psi\in C^\infty_0\big(I \times
\{a<|x|<b\}\big)$ spherical  coordinates $x=r\omega$ yields
$$
\big\langle
v(t,|x|, x/|x|)\,,\,\psi
\big\rangle
\ =\
\int_I
\int_a^b
\int_{\S^2} v(t,r,\omega)\
\psi(t,r\omega)\,r^{d-1}\,dt\,dr\,d\omega\,.
$$
Since the map
$$
C^\infty_0\big(I \times
\{a<|x|<b\}\big)
\ \ni\
\psi
\quad
\mapsto
\quad
\psi(t,r\omega)\,r^{d-1}
\ \in\
C^\infty_0(I\times ]a,b[\times \S^2)
$$
is continuous, it follows that $v(t,|x|,x/|x|)$
is a well defined distribution on
$I \times
\{a<|x|<b\}$ for any distribution $v\in {\mathcal D}^\prime(I\times]a,b[\times \S^2)$. 
Our first step concerns the wave equation.

\begin{theorem}
\label{thm:nogo2}
Suppose that $v\in {\mathcal D}^\prime(\{I\times]a,b[\times \S^2\})$
is a radial solution of the wave equation as a function of $(t,x)$, that is
$$
\Box_{t,x}v(t,|x|,\omega)\ =\ 0
\quad
{\rm in \ the\ sense \ of\ distributions\ on}\ I\times \{a<|x|<b\}\times\S^2.
$$
Then
the distribution $v(t,|x|,x/|x|)$ satisfies the wave equation
on $I\times \{a<|x|<b\}$ if and only if $v$ is independent of 
$\omega$. 
\end{theorem}

\noindent
\begin{proof}  If $f(\omega)$ is a distribution
on $\S^2$ then the distribution $f(x/|x|)$ on 
$\{x\ne 0\}$ has vanishing radial derivatives
so
$$
\Delta_x \big(f(x/|x|)\big)
\  =\
 |x|^{-2}\Delta_\omega f\big|_{\omega=x/|x|}\,,
$$
where $\Delta_\omega$ denotes the Laplace
Beltrami operator on $\S^2$.
Therefore
\begin{equation}
\Box_{t,x}\big(v(t,|x|,x/|x|)\big)
\ =\ 
\Box_{t,x}v(t,|x|,\omega)\big|_{\omega=x/|x|}
\ -\ 
|x|^{-2}\Delta_\omega v\big|_{\omega=x/|x|}\,.
\end{equation}
By hypothesis the first term vanishes and one
concludes that
$
\label{eq:harmonic}
\Delta_\omega v\big|_{\omega=x/|x|}
=
0$.
Thus $v(t,r,\cdot)$  is harmonic 
as a function of $\omega\in S^2$ so is 
independent of $\omega$.
\end{proof}

\begin{theorem}
\label{thm:nogo3}
Suppose that
$w\in {\mathcal D}^\prime(I\times[a,b[\times \S^2\,:\,\CC^6)$
satisfies 
\begin{equation}
\label{eq:boxtx}
\Box_{t,x} w(t,|x|,\omega) =0\,,
\quad
{\rm in}
\quad
I\times \{a<|x|<b\}\times \S^2
\,.
\end{equation}
Then
$u:=w(t,|x|,x/|x|):=\big(E(t,x),B(t,x)\big)$ satisfies
$$
\Box u=0
\qquad
{\rm and}
\qquad \dive E\ =\ \dive B\ =\ 0
\quad{\rm in}\quad
 I\times  \{a<|x|<b\},
 $$
 if and only if
there are constant vectors $c_1$ and $c_2$ so that
 $u=c_1 + c_2t$.
\end{theorem}

\noindent
\begin{proof}  Theorem \ref{thm:nogo2} implies that $w$ is 
independent of $\omega$.  
Choose $\rho\in C^\infty_0(]0,1[)$
with $\int \rho(s)\ ds =1$.
For $1>>\eps>0$ define
$$
w^\eps \ :=\ 
\int w(t+\eps s, r)\ \rho(s) \ ds
\qquad
{\rm and}
\qquad
u^\eps := w^\eps(t,|x|)\,.
$$ 
Since $u$ is the limit of the $u^\eps$ in the sense
of distributions,
it suffices to prove that $w^\eps$ is of the form  $c_1^\eps +c_2^\eps t$.

Since $w^\eps$ is smoothed in $t$, its wavefront set is contained
in $\{(t,x,\tau,\xi)\ :\ \tau=0\}$.  Since it is a solution of the wave
equation, the wavefront set is a subset of the characteristic
variety $\{\tau^2=|\xi|^2\}$.  Since these sets are disjoint in
$\{(\tau,\xi)\in \R^{1+3}\setminus 0\}$, $w^\eps\in C^\infty$.
Thus it is sufficient to prove the result for smooth $w$ independent
of $\omega$.

For such $w$ compute
$$
\dive E(t,|x|)
\ =\ 
\sum_{j=1}^3
\frac{\partial E_j}{\partial r}\
\frac{\partial r}{\partial x_j}
\quad
{\rm for}
\quad
a<|x|<b\,.
$$
Use $\dive E=0$ and multiply by $2r$ to find
$$
0
\ =\
\sum_{j=1}^3
\frac{\partial E_j}{\partial r}\
\frac{\partial r^2}{\partial x_j}
\ =\ 
\sum_{j=1}^3
\frac{\partial E_j(t,r)}{\partial r}\
2x_j
\,.
$$
If $E$ is not independent of $r$, 
choosing $x$ parallel to $\partial_r E$
yields a contradiction.  
Therefore $E$ is independent of $r$.  The 
same argument shows that $B$ is independent of $r$.

Since $\big(E(t),B(t)\big)$ satisfies the wave equation
it must be a linear function of $t$.  
\end{proof}

The next Corollary shows that they are no solutions
of the dynamic equations alone, that is without the
divergence equations \eqref{eq:1.2}.

\begin{corollary}  If
$w\in\caD^\prime(I\times]a,b[\times \S^2)$
satisfies \eqref{eq:boxtx}
and
$u:=w(t,|x|,x/|x|)$ satisfies
the dynamic Maxwell's equations \eqref{eq:1.1},
then there is a constant $c$ and a distribution $f\in \caD^\prime(\{a<|x|<b\})$ so that $u=f(x) +c\,t$.
 \end{corollary}

\noindent
\begin{proof}  Denote $u=\big(E(t,x),B(t,x)\big)$.
The function $v=\partial_tu = \big(\curl B\,,\,-\curl E\big)$
is then of the  form $\tilde w(t,|x|,x/|x|)$ and  satisfies 
the dynamic equations
{\sl and} the divergence conditions
\eqref{eq:1.2}.  Theorem \ref{thm:nogo3}
implies that $v=v(t)$ is a linear function of $t$. Plugging
this into the dynamic equation, one finds that $v$ must be constant.

Therefore $u_t$ is constant.
It follows that
$
u = 
c\,t +  f(x),
$
with $f$ a distribution depending only on $x$.
\end{proof}

\section{Incoming solutions of Maxwell's equations.}

\begin{theorem}
\label{thm:incoming}
If $h\in C^\infty(\R)$ and for all $k$, $ \partial^kh(s)\in L^1([0,\infty[)$,
then,
\begin{equation}
\label{eq:E}
E \ :=\ 
\bigg(
\frac{h^{\prime\prime}(|x|+t)}{|x|}
\ -\
\frac
{h^\prime(|x|+t)}{|x|^2}
\bigg)\,
\frac{x}{|x|}
\,\wedge\,
(1,0,0)\,,
\end{equation}
\begin{equation}
\label{eq:B}
B\ :=\
-\bigg(
\frac{h^{\prime\prime}}{|x|}
-
\frac{3\,h^{\prime}}{|x|^2}
+
\frac{3\,h}{|x|^3}
\bigg)\ 
\frac{x}{|x|}
\wedge\bigg(
\frac{x}{|x|}
\wedge
\big(1,0,0\big)
\bigg)
\ +\
2\,\bigg(
\frac{h^{\prime}}{|x|^2}
-
\frac{h}{|x|^3}
\bigg)\,
\big(1,0,0\big),
\end{equation}
where
the argument of the functions $h^{(k)}$ 
is $|x|+t$,
define smooth divergence free incoming solutions
of Maxwell's equations in $\R_t \times (\R^3 \setminus  0)$.
 \end{theorem}

\noindent
{\bf Proof.}
Since
$$
g(t,x)\ :=\ 
\frac{f(|x| + t)}{|x|}
$$ 
is an incoming spherical solution of the wave equation,
$$
E\ =\ \curl (g,0,0)
\ = \
\left |
\begin{matrix}
i & j& k
\cr
\partial_1&\partial_2&\partial_3
\cr
g&  0 & 0
\end{matrix}
\right |
\ =\
(0, \partial_3g,-\partial_2 g)
$$
 is a divergence free incoming solution 
of the
vector wave equation.

The derivatives $\partial_jg$ have two terms from the product rule,
\begin{equation}
\label{eq:firstgderivs}
\partial_j g
\ =\ 
\frac{f^\prime(|x|+t)\ \partial_j|x|}{|x|}
\ -\
\frac
{f(|x|+t)\ \partial_j|x|}{|x|^2}\,,
\qquad
\partial_j|x|\ =\ 
\frac{x_j}{|x|}\,.
\end{equation}
Therefore,
\begin{equation}
\label{eq:Evectorial}
E\ =\ 
\bigg(
\frac{f^\prime(|x|+t)}{|x|}
\ -\
\frac
{f(|x|+t)}{|x|^2}
\bigg)\,
\frac{x}{|x|}
\,\wedge\,
(1,0,0)\,.
\end{equation}
The vector field $(x/|x| )\,\wedge\,(1,0,0)$ on the sphere $|x|=1$
winds around the latitude lines  which are the intersections
with the planes $x_1={\rm const.}$

Compute
\begin{equation}
\label{eq:dtdxg}
\partial_t E\ =\ 
\Big(
0\,,\,
\partial_t\partial_3g\,,\,
-\partial_t\partial_2g
\Big)\,,
\quad\ \
\partial_t\partial_j g
\ =\ 
\frac{f^{\prime\prime}(|x|+t)\ \partial_j|x|}{|x|}
\ -\
\frac
{f^\prime(|x|+t)\ \partial_j|x|}{|x|^2}\,,
\end{equation}
to find
\begin{equation}
\label{eq:etvectorial}
\partial_tE\ =\ 
\bigg(
\frac{f^{\prime\prime}(|x|+t)}{|x|}
\ -\
\frac
{f^\prime(|x|+t)}{|x|^2}
\bigg)\,
\frac{x}{|x|}
\,\wedge\,
(1,0,0)\,.
\end{equation}

Since
$
x\wedge (1,0,0)=(0,x_3,-x_2)$,
one has
\begin{equation}
\label{eq:xwedgex}
x\wedge\big(x\wedge (1,0,0)\big)
\ =\ 
\left|
\begin{matrix}
i & j& k\\
x_1&x_2&x_3\\
0 & x_3&-x_2
\end{matrix}
\right|
\ =\
\Big(-x_2^2 -x_3^2\,,\, x_1x_2\,,\, x_1x_3\Big)
.
\end{equation}

Write 
 $$
\curl E = \curl
\Big(\phi
\ 
\frac{x}{|x|}
\,\wedge\,
(1,0,0)\Big)
\,,
\qquad
{\rm with }
\qquad
\phi:=
\bigg(
\frac{f^\prime(|x|+t)}{|x|}
\ -\
\frac
{f(|x|+t)}{|x|^2}
\bigg).
$$
Use the product rule
\begin{equation}
\label{eq:product}
\curl (\phi\, \bfw)= (\grad \phi)\wedge \bfw +\phi\,\curl \bfw 
\end{equation}
to find
\begin{equation}
\label{eq:curlE}
\curl E =(\grad g)
\wedge \Big(
\frac{x}{|x|}\,\wedge (1,0,0)\Big)
 + 
g\,
\curl\Big(
\frac{x}{|x|}\,\wedge(1,0,0)
\Big).
\end{equation}

Next evaluate the
curl term in \eqref{eq:curlE}. 
Identity \eqref{eq:product} yields
\begin{equation}
\label{eq:curl1}
\curl
\bigg(
\frac{1}{|x|}
\big(0\,,\,
x_3\,,\,-x_2\big)
\bigg)
=
\grad
\bigg(
\frac{1}{|x|}\bigg)
\wedge
\big(0\,,\,
x_3\,,\,-x_2\big)
+
\frac{1}{|x|}\,\curl
\big(0\,,\,
x_3\,,\,-x_2\big)\,.
\end{equation}
Since
$\grad |x|^{-1} = - |x|^{-2}\, x/|x|$,
the first term in \eqref{eq:curl1}
is equal to
\begin{equation}
\label{eq:curl2}
\grad
\bigg(
\frac{1}{|x|}\bigg)
\wedge
\big(
x\wedge (1,0,0)\big)
=
-\,\frac{1}{|x|^3}\,
\Big(
x\wedge\big(x\wedge (1,0,0)\big)\Big)
 =
-\,\frac{1}{|x|}\
 \frac{x}{|x|}\wedge\bigg(
\frac{x}{|x|}\,\wedge(1,0,0)\bigg)
.
\end{equation}

For the second term in \eqref{eq:curl1}, use
\begin{equation}
\curl (0,x_3,-x_2)
\ =\
\left|
\begin{matrix}
i & j & k\cr
\partial_1 &\partial_2& \partial_3\cr
0 & x_3 & -x_2
\end{matrix}
\right|
\ =\
(-2\,,\,0\,,\,0)\,.
\end{equation}
Adding 
yields
\begin{equation}
\label{eq:curlfriend2}
\curl\Big(
\frac{x}{|x|}\,\wedge(1,0,0)
\Big)
\ =\
-\,\frac{1}{|x|}\
 \frac{x}{|x|}\wedge\bigg(
\frac{x}{|x|}\,\wedge(1,0,0)\bigg)
\ -\ \frac{2}{|x|}\\(1,0,0)\,,
\end{equation}
\begin{equation}
\label{eq:half}
\phi\,
\curl\Big(
\frac{x}{|x|}\,\wedge(1,0,0)
\Big)
=
\bigg(-\frac{f^\prime}{|x|^2}
+
\frac{f}{|x|^3}
\bigg)
 \frac{x}{|x|}\wedge\bigg(
\frac{x}{|x|}\,\wedge(1,0,0)\bigg)
-
\frac{2\phi}{|x|}(1,0,0).
\end{equation}

The field 
\begin{equation}
\label{eq:longitude}
\frac{x}{|x|}
\ \wedge\
\bigg(\frac{x}{|x|}
\ \wedge\
(1,0,0)\bigg)
\  = \ 
-\,
\big(1,0,0\big)_{tan} 
\end{equation}
is equal to the tangential part of $(-1,0,0)$
on the sphere.  It flows from the pole $x_1=-1$
to the opposite pole along the longitude lines.

The first summand in \eqref{eq:curlE} is equal to
\begin{equation}
\label{eq:gradgterm}
\grad \phi\wedge\bigg(\frac{x}{|x|}\big(1,0,0\big)\bigg)
\ =\ 
\partial_r \phi\  \frac{x}{|x|}\wedge\bigg(
\frac{x}{|x|}\,\wedge(1,0,0)\bigg)\,.
\end{equation}
The derivative $\partial_r\phi$ 
has terms  where the derivative
hits the $f$ factor and those when the derivative
hits the $|x|^{-p}$.  This yields
\begin{equation}
\label{eq:gradg}
\partial_r \phi
\ =\ 
\frac{f^{\prime\prime}}{|x|}
-\frac{f^\prime}{|x|^2}
-\frac{f^\prime}{|x|^2}
+\frac{2\,f}{|x|^3}
\ =\
\frac{f^{\prime\prime}}{|x|}
-\frac{2\,f^\prime}{|x|^2}
+\frac{2\,f}{|x|^3}
\,.
\end{equation}
So,
\begin{equation}
\label{eq:gradg2}
\grad \phi\wedge\bigg(\frac{x}{|x|}\big(1,0,0\big)\bigg)
\ =\ 
\bigg(
\frac{f^{\prime\prime}}{|x|}
-\frac{2\,f^\prime}{|x|^2}
+\frac{2\,f}{|x|^3}
\bigg)
 \frac{x}{|x|}\wedge\bigg(
\frac{x}{|x|}\,\wedge(1,0,0)\bigg)\,.
\end{equation}
Summing
 \eqref{eq:half}
and \eqref{eq:gradg2} yields
\begin{equation}
\label{eq:curlE2}
\curl E
\ =\ 
\bigg(
\frac{f^{\prime\prime}}{|x|}
-
\frac{3\,f^\prime}{|x|^2}
+
\frac{3\,f}{|x|^3}
\bigg)\ 
\frac{x}{|x|}
\wedge\bigg(
\frac{x}{|x|}
\wedge
\big(1,0,0\big)
\bigg)
\ -\
2\,\bigg(
\frac{f^{\prime}}{|x|^2}
-
\frac{f}{|x|^3}
\bigg)\,
\big(1,0,0\big)
\,. 
\end{equation}

Take $h\in \cap_sW^{s,1}([0,\infty[)$ as in the Theorem
 and set $f=h^\prime$.
Then the equation
$B_t = -\curl E$ 
can be integrated using \eqref{eq:curlE2},
\begin{align*}
B\ &=\
-\int_t^\infty
-\curl E(t,x)\, dt
\cr
&=\
\int_t^\infty
\bigg(
\frac{f^{\prime\prime}}{|x|}
-
\frac{3\,f^\prime}{|x|^2}
+
\frac{3\,f}{|x|^3}
\bigg)\ 
\frac{x}{|x|}
\wedge\bigg(
\frac{x}{|x|}
\wedge
\big(1,0,0\big)
\bigg)
\ -\
2\,\bigg(
\frac{f^{\prime}}{|x|^2}
-
\frac{f}{|x|^3}
\bigg)\,
\big(1,0,0\big)\ \,dt\cr
&=\
\int_t^\infty
\bigg(
\frac{h^{\prime\prime\prime}}{|x|}
-
\frac{3\,h^{\prime\prime}}{|x|^2}
+
\frac{3\,h^\prime}{|x|^3}
\bigg)\ 
\frac{x}{|x|}
\wedge\bigg(
\frac{x}{|x|}
\wedge
\big(1,0,0\big)
\bigg)
\ -\
2\,\bigg(
\frac{h^{\prime\prime}}{|x|^2}
-
\frac{h^\prime}{|x|^3}
\bigg)\,
\big(1,0,0\big)
\ \,dt
\cr
&=\
-\bigg(
\frac{h^{\prime\prime}}{|x|}
-
\frac{3\,h^{\prime}}{|x|^2}
+
\frac{3\,h}{|x|^3}
\bigg)\ 
\frac{x}{|x|}
\wedge\bigg(
\frac{x}{|x|}
\wedge
\big(1,0,0\big)
\bigg)
\ +\
2\,\bigg(
\frac{h^{\prime}}{|x|^2}
-
\frac{h}{|x|^3}
\bigg)\,
\big(1,0,0\big)\,.
\end{align*}

 By construction, $E$ and $B$ are
 divergence free, $\Box E=0$,  and
 $B_t=-\curl E$.  We must prove that $E_t-\curl B=0$.  Using
 the fact that $\dive E=0$, compute
 $$
 \partial_t\big(
 E_t-\curl B
 \big)
 \ =\ 
 E_{tt}-\curl B_t
 \ =\
 E_{tt}+\curl\curl E
 \ =\ E_{tt}-\Delta E\ =\ 0\,.
 $$
 So,
 for each $x$, $E_t(t,x)-\curl B(t,x)$ 
 is independent of $t$.
The formulas for $E,B$  show that  it tends to zero as $t\to \infty$.
Therefore it vanishes identically.
  This completes the proof of the Theorem.
\qed

\section{Disappearing solutions for Maxwell's Equations.}

We construct asymptotically disappearing solutions 
in $|x|>1$ that satisfy
a homogeneous boundary condition
\begin{equation}
\label{eq:bc}
u=(E,B)\  \in \ \caN(x),\quad
{\rm on}\quad |x|=1\,.
\end{equation}
Here $\caN(x)$ is a four dimensional linear subspace
of $\CC^6$ depending smoothly on $x$.   Therefore,
\eqref{eq:bc} represents two linear
constraints on the  boundary values.

Write the Maxwell equations in matrix form
$$
u_t 
\ +\ 
\sum_{j=1}^3
A_j\,
\partial_j u
\ =\ 
0\,.
$$
The $A_j$ are real symmetric matrices and for $0\ne \xi\in \R^3$,
$A(\xi) := \sum A_j\xi_j$ has rank equal to 4. Denote by
$n(x)$ the outward unit normal to the boundary.
For solutions of Maxwell's equations 
in $|x| > 0$ suitably smooth and small at infinity,
one has
\begin{equation}
\label{eq:decay}
\frac{d}{dt}\,
\int_{|x|>1}
|u|^2\ dx
\ =\ \int_{|x|=1} \big\langle
A(n(x))u\,,\, u\big\rangle
\ d\sigma\,.
\end{equation}
The theory of maximal
dissipative boundary value problems of 
Friedrichs \cite{F} as extended by Lax and Phillips \cite{LP}
shows that a sufficient condition for $\caN$ to define a well
posed mixed initial boundary value problem generating a 
contraction semigroup on $L^2(\{|x|\ge 1\})$ is that
$$
{\rm dim}\, \caN=4,
\quad
{\rm and},
\quad
\forall |x|=1,\ \ 
\forall u\in \caN(x),\ \ 
\big\langle
A(n(x)) u\,,\, u\big\rangle \ \le \ 0.
$$
Such spaces are called {\bf maximal dissipative}.  It follows
that $\caN(x)\supset {\rm Ker}\,A(n(x)$ for all boundary points
$x$.

For any unit vector $n$ the eigenvalues of $A(n)$ are
$-1,0,1$.  The kernel is the set of $(E,B)$ with both
$E$ and $B$ parallel to $n$.   The condition that
$\caN$ contain the kernel is equivalent to saying that
belonging to $\caN$ is determined entirely by the tangential
components $(E_{tan},B_{tan})$.  The eigenspace corresponding
to eigenvalue $\pm 1$ is equal to 
$$
E_\pm 
\ :=\ 
\big\{
(E,B)\,:\,
E_{tan}\ =\ \pm\, n\wedge B_{tan}
\big\}\,.
$$
The span of eigenspaces $E_-\oplus {\rm ker}\, A(n)$ with
nonpositive eigenvalues   
satisfies the strict dissipativity identity
 $$\
 \forall u\in E_-\oplus {\rm Ker}\,A(n),
 \quad
 \big\langle
 A(n) u\,,\, u\big\rangle 
\ =\ 
-
\|u_{tan}\|^2
\ =\ 
 - \big\| E_{tan}, B_{tan} \big\|^2\,.
 $$
 The vector $(E_{tan}, B_{tan} )$ is the projection
 orthogonal to ${\rm Ker}\, A_n$. 
The 
Poynting vector is $E\wedge B$ safisfies  $\langle A(n) u\,,\,u\rangle = (E\wedge B) \cdot n$.

\begin{lemma}
\label{lem:hresidue}  The solutions of Theorem \ref{thm:incoming}
satisfy on each sphere
\begin{equation}
\label{eq:lemma4.1}
E_{tan} \ -\
n\wedge B_{tan} 
\ =\ 
-\,\frac{h}{|x|^3}
\,
\Big(
\frac{x}{|x|}\ \wedge\  (1,0,0)
\Big)\,.
\end{equation}
\end{lemma}

\noindent
{\bf Proof.}  On spheres centered at the origin
 the  solutions from Theorem \ref{thm:incoming} 
 satisfy $E=E_{tan}$.
For all fields, 
$n\wedge B_{tan}= n\wedge B$.  By using \eqref{eq:longitude},
we find
$$
\begin{aligned}
n\wedge B_{tan}
\ &=\ 
\bigg(
\frac{h^{\prime\prime}}{|x|}
-
\frac{3\,h^{\prime}}{|x|^2}
+
\frac{3\,h}{|x|^3}
\bigg)\ 
\bigg(
\frac{x}{|x|}
\wedge
\big(1,0,0\big)
\bigg)
\ +\ 
2\,\bigg(
\frac{h^{\prime}}{|x|^2}
-
\frac{h}{|x|^3}
\bigg)\,
\bigg(
\frac{x}{|x|}
 \wedge
(1,0,0)
\bigg)\cr
\ &=\
\bigg(
\frac{h^{\prime\prime}}{|x|}
-
\frac{h^{\prime}}{|x|^2}
+
\frac{h}{|x|^3}
\bigg)
\bigg(
\frac{x}{|x|}
 \wedge
(1,0,0)
\bigg)\,.
\end{aligned}
$$
Subtracting from $E$  completes the proof.
\qed

 \begin{lemma}
 There is an $\eps_0>0$ so that for  $|\eps|<\eps_0$ the boundary space
 $$
 \caN_\eps
 \ :=\ 
 \big\{ (E,B)\, :\,
(1 + \eps) E_{tan} \ =\  n\wedge B_{tan}
\big\}
\ \supset\
{\rm Ker}\,A_n
$$
is maximal dissipative.  On $|x|=1$ it satisfies,
 with a constant $c>0$ independent
of $\eps_0, u$,
$$\
 \forall u\in \caN_\eps,
 \quad
 \big\langle
 A(n) u\,,\, u\big\rangle 
\  \le\
-\,c\,\|u_{tan}\|^2
\ =\ 
 -\,c\, \big\| (E_{tan}, B_{tan}) \big\|^2\,.
 $$
 \end{lemma}
  
  \noindent
{\bf Proof.}  Write 
$$
 \caN_\eps\ \ni\
u\ = \
 (n\wedge B_{tan}
 \,,\,
  B_{tan}) 
 \ -\
 (\eps E_{tan}\,,\,0)\ :=\ v\ +\ w.
$$
$$
\big\langle
A(n) u
\,,\,
u
\big\rangle
\ =\
\big\langle
A(n) v
\,,\,
v
\big\rangle
\  +\ 
2
\big\langle
A(n) v
\,,\,
w
\big\rangle
\ +\ 
\big\langle
A(n) w
\,,\,
w
\big\rangle\,.
$$
Let $v=(\widetilde E, \widetilde B)$.  By definition
$v\in \caN_0$ and   the strict dissipativity of $\caN_0$ implies
$$
\begin{aligned}
\big\langle
A(n) v
\,,\,
v
\big\rangle
\ &\le\
-\,
\|v_{tan}\|^2
\ =\ 
-\|\widetilde E_{tan}\|^2 
-\|\widetilde B_{tan}\|^2 \cr
\ &=\ 
-\|n\wedge B_{tan}\|^2 -\|B_{tan}\|^2
\ =\ 
-\|(1+\eps)^{-1}E_{tan}\|^2 
-\|B_{tan}\|^2 
\,.
\end{aligned}
$$
Next estimate
$$
\big|\big\langle
A(n) w
\,,\,
w
\big\rangle
\big|
\ \le\
\|w_{tan}\|^2
 \ =\ 
 \,\|\eps\,E_{tan}\|^2
 \,,
$$
$$
2\big|\big\langle
A(n) v
\,,\,
w
\big\rangle
\big|
\ \le\
2\|v_{tan}\|
\,
\|w_{tan}\|
\ \le\
\frac{1}{4}\,\|v_{tan}\|^2
\ +\ 
4\,\|w_{tan}\|^2.
$$
Therefore
$$
\begin{aligned}
\big\langle
A_n u
\,,\,
u
\big\rangle
\ &\le\
-
\frac{3}{4}\,
\|v_{tan}\|^2
\ +\ 
5\,\|w_{tan}\|^2\cr
\ &=\ 
-\,\frac{3}{4}\,
\big(
\|(1+\eps)^{-1}E_{tan}\|^2 +
\|B_{tan}\|^2 
\big)
\ +\ 
5\,\|\eps\,E_{tan}\|^2\,.
\end{aligned}
$$
and the proof is complete.
\qed

\vskip.2cm

Combining \eqref{eq:E} with  Lemma \ref{lem:hresidue},   
shows that  the solutions of Theorem \ref{thm:incoming} satisfy
\begin{equation}
\label{eq:residue}
( 1 + \eps) E_{tan} \ -\  n\wedge B_{tan}
\ =\ 
\bigg(
\frac{\eps h^{\prime\prime}}
{|x|}
\ -\ 
\frac{\eps h^\prime}
{|x|^2}
\ - \
\frac{h}
{|x|^3}
\bigg)
\Big(
\frac{x}{|x|}
\,\wedge\,
(1,0,0)\Big)
\,.
\end{equation}
Asymptotically disappearing solutions are constructed by choosing $h$
so that the right hand side vanishes.

\begin{theorem} [Asymptotically disappearing Maxwell solution.]  
\label{thm:asympt} 
With 
$\eps_0$ from the preceding
 Lemma and $0<\eps<\eps_0$, define 
$2r:= 1- \sqrt{1 + 4/\eps}<0$ and $h(s):=e^{rs}$.  Then $(E,B)$ defined by 
\eqref{eq:E} and \eqref{eq:B} yield a
divergence free solution
of the maximal dissipative boundary value problem defined
by the Maxwell equations in $|x|>1$ with 
maximal dissipative boundary condition
$$
(1 + \eps) E_{tan} \ -\  n\wedge B_{tan}
\ =\ 
0,
\qquad
{\rm on}
\qquad
|x|=1\,.
$$
For each $\alpha$ 
there is a constant $C(\eps, \alpha)$ so that
$
\big|\partial^\alpha (E,B)
\big|
\le
C\, h(t+|x|)$.  In particular, the energy decays exponentially
as $t\to \infty$.
\end{theorem}

\noindent
{\bf Proof.}   
The solutions of 
 the ordinary differential equation
$\eps h^{\prime\prime} -\eps h^\prime - h\ =\ 0$
 are $e^{\rho s}$ with  $\rho$ satisfying 
$\eps \rho^2 -\eps \rho -1=0$.  The roots of the last equation are
$$
\rho\ =\ 
\frac{\eps \pm \sqrt{\eps^2 + 4\eps}}
{2\eps}
\ =\ 
\frac{1}{2}
\Big(
1\ \pm\
\sqrt{1 + 4/\eps}
\Big)
\,.
$$
The  sign $-$  yields $r$ and it follows that $h=e^{rs}$ is a solution.
Equation \eqref{eq:residue} shows that the boundary
condition is satisfied.  The estimates are immediate
consequences of the formulas.
\qed

\begin{remark}   The semigroup of contractions
on $L^2$ defined by the maximal dissipative boundary value
problem is of the form $e^{tG}$ with generator $G$
(see for instance, Ch. $III$ in \cite{P}).
The solutions in Theorem \ref{thm:asympt} have the form
$e^{rt}\phi(x)$  with  $G\phi=r\phi$.
\end{remark}

The preceding strategy yields profiles which
decay exponentially and never have compact support.
It yields only asymptotically disappearing
solutions.
The next computations show that 
it is possible to choose $h$ with compact support 
so that the resulting function 
satisfies a {\it time dependent}
  boundary condition.
Let
\def\eb{\exp\Bigl(-\,\frac{1}{b^2 - y^2}\Bigr)}
\begin{equation}
 \label{eq:4.4}
h(y) \ :=\
 \begin{cases}
 \eb,\quad|y| < b,
 \\
  0,\hskip2.4cm  |y| \geq b > 1 
 \end{cases}
  ,
   \qquad
   h\in C^\infty_0(\RR)\,.
\end{equation}

\begin{theorem} [Disappearing Maxwell solutions]
\label{thm:disapp}  For any $\delta\in ]0,1[$ 
and $h$ given by $(\ref{eq:4.4})$  
we can choose $b>1$
and  $\gamma(t)\in C^\infty_0(\RR)$ with  $|\gamma|< \delta$
so that  $(E,B)$ defined by  \eqref{eq:E} and \eqref{eq:B} 
is a
divergence free solution of 
Maxwell's equations in $t\ge 0\,,\,|x|\ge 1$ satisfying
the maximal strictly dissipative boundary condition
$$
(1 + \gamma(t))E_{\tn} 
\ -\ n \wedge B_{\tn} 
\ = \
0
\qquad
{\rm on}
\qquad |x| = 1.
$$
\end{theorem}

\noindent
{\bf Proof.}
Compute with $b>1$ to be chosen below,
$$
h'(y) \ =\
 \frac{-2 y}{(b^2 - y^2)^2} \eb,\qquad |y| < b,
 $$
$$
\begin{aligned}
h''(y) 
 \ &=\
  \frac{2(3y^4 - 2(b^2 -1)y^2 - b^4)}{(b^2 - y^2)^4} \eb, \qquad |y| < b.
  \end{aligned}
  $$
 For  the electric field on $|x|=1$ compute
\begin{equation}
\label{eq:bizarre}
\begin{aligned}
{h''(y)} - {h'(y)}
\ &=\
 \frac{2(3y^4 - 2(b^2 - 1)y^2 - b^4) + 2y(b^2 - y^2)^2}
 { (b^2 - y^2)^4} 
 \ \eb
 \cr
 \ &:=\
 \frac{Q(y,b)}
 { (b^2 - y^2)^4} 
  \ \eb
  \ =\ 
   \frac{Q(y,b)}
 { (b^2 - y^2)^4} 
 \ 
 h,
 \qquad
 |y|<b
 \,.
 \end{aligned}
 \end{equation}

 Since $Q(1,1)= 4$ and $Q$ is continuous,
 there is a $\mu\in ]0,1]$ so that 
 $Q(y,b)\ge 3$ on $[1,1+\mu]\times [1,1+\mu]$.

Now set $y = |x| + t,\: |x| \geq 1, \: t \geq 0$. 
 For $1 \leq 1 + t < b\le1+\mu$
 equation \eqref{eq:bizarre} asserts that
$$
{h(1+ t)}
\ =\
  \frac{(b^2 - (1 + t)^2)^4}{Q(1 + t,b)}
 \ 
 \Bigl[ {h''(1 + t)} - {h'(1 + t)}\Bigr]
 $$
and on $|x|=1$ the right hand side of \eqref{eq:lemma4.1}
satisfies
$$ 
-\,{h(1 + t)} \Bigl(\xx \wedge (1, 0, 0) \Bigr)
\ =\
-\,  \frac{(b^2 - (1 + t)^2)^4}{Q(1 + t,b)} 
 \ 
 E_{\tn} 
 \ :=\
-\,  \gamma(t) \,E_{\tn}, \qquad 1\le 1 + t \le b,
  $$
  defining $\gamma\in C^\infty([0,b-1])$.
  Taking  $b<1+\mu$ sufficiently close to $1$  
guarantees that 
$|\gamma(t)|< \delta/2$ for $0\le t\le b-1$.

For $|x| = 1$ and $1 \leq  1 + t < b$ the
solution satisfies the
maximal strictly dissipative boundary condition
\begin{equation}
\label{eq:gammabc}
0\ =\ 
\Bigl( E_{\tn} - n \wedge B_{\tn}\Bigr) 
\ +\
 {h(1 + t)} \Bigl( \xx \wedge (1, 0, 0)\Bigr)
 \ =\
 \Bigl( 1 + \gamma(t)\Bigr) E_{\tn} \ -\  \Bigl(n \wedge B_{\tn}\Bigr) 
 \,.
 \end{equation}

  Extend
 $\gamma$ to an element of $C^\infty_0(\RR)$ with
 $|\gamma|<\delta$.  
 Since the solution vanishes
 identically on $|x|=1$ for $t\ge b-1$, the
 boundary condition 
 \eqref{eq:gammabc}  is satisfied in $t\ge 0\,,\,|x|=1$.
 \qed

\vskip.3cm

\noindent
{\footnotesize

\end{document}